\documentclass{article}

\usepackage{arxiv}

\usepackage[utf8]{inputenc} 
\usepackage[T1]{fontenc}    
\usepackage{hyperref}       
\usepackage{url}            
\usepackage{booktabs}       
\usepackage{amsfonts}       
\usepackage{nicefrac}       
\usepackage{microtype}      
\usepackage{lipsum}		
\usepackage{graphicx}

\usepackage{amsmath, amssymb, amsthm}
\usepackage{enumitem}

\theoremstyle{plain}
\newtheorem{theorem}{Theorem}[section]

\theoremstyle{definition}
\newtheorem{definition}[theorem]{Definition}

\theoremstyle{remark}
\newtheorem{remark}{Remark}

\newcommand{\ol}[1]{{\overline{#1}}}

\title{A remark on statistics for detecting laboratory effects in ORDANOVA}

\date{\today} 					

\author{ \href{https://orcid.org/0000-0002-3471-1594}{\includegraphics[scale=0.06]{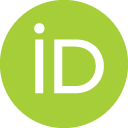}\hspace{1mm}Jun-ichi Takeshita}\\
	Research Institute of Science for Safety and Sustainability\\
	National Institute of Advanced Industrial Science and Technology (AIST) \\
	Tsukuba, Ibaraki, Japan\\
	\texttt{jun-takeshita@aist.go.jp} \\
	\And
	Yuto Arai \\
	Department of Industrial Administration, Faculty of Science and Technology\\
	Tokyo University of Science\\
	Noda, Chiba, Japan
	\AND
	Mayu Ogawa \\
	Department of Industrial Administration, Faculty of Science and Technology\\
	Tokyo University of Science\\
	Noda, Chiba, Japan
	\And
	Xiao-Nan Lu \\
        Department of Computer Science and Engineering, Faculty of Engineering\\
        University of Yamanashi\\
        Kofu, Yamanashi, Japan
	\And
	Tomomichi Suzuki\\
		Department of Industrial Administration, Faculty of Science and Technology\\
		Noda, Chiba, Japan\\
	Tokyo University of Science
}

\date{\today}

\renewcommand{\headeright}{Jun-ichi Takeshita \textit{et al.}}
\renewcommand{\undertitle}{}

\begin{document}
\maketitle

\begin{abstract}
    The present study provides a test statistic for detecting laboratory effects in the analysis of ordinal variation (ORDANOVA).
  The ORDANOVA is a similar analysis method to one-way analysis of variance for ordinal data; however, no appropriate statistics have been proposed for conducting statistical tests to detect laboratory effects.
  The present study proves that a statistic for representing an ordinal variation in ORDANOVA also works for conducting statistical tests for detecting laboratory effects for an arbitrary number of ordinal levels, especially more than three ordinal levels.
  In addition, a real-world example involving data from a collaborative study for an in vivo test is analyzed using the proposed statistic.
\end{abstract}



\section{Introduction}
Collaborative studies are conducted to evaluate accuracy of new measurement methods and results; one of the primary aims of these studies is to verify whether laboratory effects exist or not.
International Organization for Standardization (ISO) 5725 Parts 1~\cite{ISO_1994_ISO57251} and 2~\cite{ISO_1994_ISO57252} describe how to organize such studies and analyze the obtained measured values.
Specially, Part 2~\cite{ISO_1994_ISO57251} uses one-way analysis of variance (ANOVA) to detect laboratory effects for quantitative measured values.
One the other hand, the demand to analyze qualitative measured values from some collaborative studies is increasing.
For example, studies on the detection of Listeria monocytogenes in foods~\cite{Scotter.Langton.ea_2001_ValidationISO} and on new real-time polymerase chain reaction (PCR) assays for detecting transgenic rice~\cite{Grohmann.Reiting.ea_2015_CollaborativeTrial} provided us binary measured values; and a study on a screening animal testing to evaluate some effects on lungs provided us five-level ordinal measured values.

To analyze ordinal measured values, several approaches were proposed in recent years~\cite{Bashkansky.Gadrich_2011_StatisticalQuality,Gadrich.Bashkansky_2012_ORDANOVAAnalysis,Scotter.Langton.ea_2001_ValidationISO,vanWieringen.deMast_2008_MeasurmentSystem}.
Most of these approaches assumed some continuous distributions behind the obtained measured values, but in many cases we could only observe the measured values as ordinal data.
In other words, those previous studies modified some statistical methods for continuous data and directly applied them to ordinal data. Wherein, Gadrich and Bashkansky~\cite{Gadrich.Bashkansky_2012_ORDANOVAAnalysis} proposed the analysis of ordinal variation (ORDANOVA) to conduct ANOVA-like analyses for ordinal measured values.
This is a remarkable study because the ORDANOVA does not necessarily assume that the data were drawn from an underlying continuous distribution.
This only assumes a multinomial distribution to produce the obtained measured values.
They also defined a test statistic for detecting laboratory effects, which works well for two-level ordinal measured values, that is, binomial measured values; however, it dose not work for more than three levels.
The aim of the present study is to prove that one of their statistics can be used to detect laboratory effects for an arbitrary number of ordinal levels, especially for more than three levels. Our results would contribute to making the ORDANOVA more powerful and practicable.
The present study, indeed, applies the statistic to a real-world example, which is the obtained measured values of a collaborative study involving intratracheal administration testing reported in~\cite{AIST_2018_AnnualReport}; these results comprise ordinal data with five levels.

\section{Preliminaries}

This section briefly introduces the ORDANOVA; see also Bashkansky et al.~\cite{Bashkansky.Gadrich_2011_StatisticalQuality,Gadrich.Bashkansky_2012_ORDANOVAAnalysis} for details.
Let $M$ be the number of laboratories participating in a collaborative study and $K$ be the number of ordinal-level categories used in the study.
Each laboratory is coded by an integer $m \in \{1, \ldots, M\}$ and has $n$ measured values.
Note that the number of category $K$ and measured values $n$ are assumed to be identical for all the $M$ laboratories in the present study.
In addition, each category is coded by an integer $k \in \{1,\ldots,K\}$, for expressing each obtained measured value.
The order of the categories follows the numbering, that is, $1<2< \cdots <K-1<K$.
The following notation is used:

\begin{itemize}
\item $n_{km}$: number of measured values belonging to the $k$th category at laboratory $m$.
  Note that $n=\sum_{k=1}^{K}n_{km}$ holds for any $m$. 
\item $p_{km}$: probability of measured values belonging to the $k$th category at laboratory $m$.
\item $\hat{p}_{km}$: proportion of measured values belonging to the $k$th category at laboratory $m$, which is an estimate of $p_{km}$. In other words, $\hat{p}_{km}=n_{km}/n$.
\item $\bar{p}_k$: mean of $p_{km}$ with respect to $m$.
\item $\hat{\bar{p}}_k$: mean of $\hat{p}_{km}$ with respect to $m$, which is an unbiased estimate of $\bar{p}_k$.
\item $F_{km}$: cumulative probability of measured values belonging up to the $k$th category at laboratory $m$.
  In other words, $F_{km}=\sum_{i=1}^{k}p_{im}$.
\item $\hat{F}_{km}$: cumulative proportion of the measured values belonging up to the $k$th category at laboratory $m$, which is an unbiased estimate of $F_{km}$.
  In other words, $\hat{F}_{km}=\sum_{i=1}^{k} \hat{p}_{im}$.
\item $\bar{F}_k$: mean of $F_{km}$ with respect to $m$.
  \item $\hat{\bar{F}}_k$: mean of $\hat{F}_{km}$ with respect to $m$, which is an unbiased estimate of $\ol{F}_k$.
\end{itemize}

Gadrich and Bashkansky~\cite{Gadrich.Bashkansky_2012_ORDANOVAAnalysis} have defined the total ordinal variation in interlaboratory comparison,
\begin{equation}
  \label{eq:1}
  \hat{h}^2_{(T)} = \frac{1}{(K-1)/4} \sum_{k=1}^{K-1} \hat{\bar{F}}_k \left( 1 - \hat{\bar{F}}_k\right)
\end{equation}
as well as the ordinal within-laboratory variation for laboratory $m$,
\begin{equation}
  \label{eq:2}
  \hat{h}^2_{m(W)} = \frac{1}{(K-1)/4} \sum_{k=1}^{K-1} \hat{F}_{km} \left( 1 - \hat{F}_{km} \right),
\end{equation}
and the classic variation of the cumulative frequencies up to the $k$th category between laboratories,
\begin{equation}
  \label{eq:3}
  \hat{S}^2_{k(B)} = \frac{1}{M} \sum_{m=1}^M \left( \hat{F}_{km} - \hat{\bar{F}}_k \right)^2.
\end{equation}
where the capital letters $T$, $W$,  and $B$ in the parentheses mean \emph{Total}, \emph{Within}, and \emph{Between}, respectively.
It should be noted that the definitions shown in equations \eqref{eq:1} and \eqref{eq:2} were based on the statistic of ordinal variation proposed by Blair and Lacy~\cite{Blair.Lacy_2000_StatisticsOrdinal}.

Under a homogeneous condition with respect to laboratories, $p_{k1} = p_{k2} = \cdots = p_{kM} \ (\forall k \in \{1, \ldots K\})$,
theses studies introduced a relationship among \eqref{eq:1}, \eqref{eq:2}, and \eqref{eq:3}:
\begin{equation*}
  \hat{h}^2_{(T)} = \hat{h}^2_{(W)} + \hat{S}^2_{(B)} =
  \frac{1}{M} \sum_{m=1}^M \hat{h}^2_{m(W)} + \frac{1}{(K-1) / 4} \sum_{k=1}^{K-1} \hat{S}^2_{k(B)},
\end{equation*}
where $\hat{h}^2_{(W)}$ represents the average of all within-laboratory ordinal variations, while $\hat{S}_{(B)}^2$ measures the average of all between-laboratory variations.

In addition, these studies defined the following test statistic $I_{(P)}$ for detecting laboratory effects:
\begin{equation*}
  I_{(P)} = \frac{\hat{S}^2_{(B)} / df_B}{\hat{h}^2_{(T)} / df_T},
\end{equation*}
where $df_B$ and $df_T$ denote the degrees of ..., respectively, that is, $df_B = M-1$ and $df_T = N-1$.

Finally, under the homogeneous condition, $H_0: p_{k1} = p_{k2} = \cdots =p_{kM} \ (\forall k=\{1, \ldots, K\})$,
these studies proved that the test statistic $I_{(P)}$ approximately follows $\chi^2 / (M-1)$ when the number of categories is $K=2$.
Regarding the case where $K \geq 3$, the following criteria (i)-(iii) were proposed for detecting laboratory effects; 
however, no approximate distribution of $I_{(P)}$  has been verified for $K \geq 3$.
\begin{enumerate}[label=(\roman*)]
\item When $I_{(P)} > 3$, reject $H_0$.
\item When $I_{(P)} \leq 3$, accept $H_0$.
\item $1 < I_{(P)} \leq 3$  is a region of doubt.
\end{enumerate}

\section{Main results}
\label{sec:main}

\subsection{Definition of a test statistic}

\begin{definition}
  Using $\hat{h}^2_{(W)}$ and $\hat{S}^2_{(B)}$, a test statistic $I_{(N)}$ for detecting laboratory effects is defined as follows:
  \begin{equation}
    \label{eq:I(N)}
    I_{(N)}:= \hat{h}^2_{(W)} + \hat{S}^2_{(B)} + \frac{1}{(K-1)/4}\sum_{k=1}^{K-1} \left(\hat{\bar{F}}_k\right)^2.
  \end{equation}
\end{definition}

\subsection{Main theorem}

This subsection proves an approximate distribution of the statistic $I_{(N)}$ for arbitrary number of categories.
The result implies that the statistic $I_{(N)}$ can be used for detecting laboratory effects for the ORDANOVA; see Remark~\ref{rem:howtest}.

\begin{theorem}
\label{thm:main}
  Under the homogeneous condition, $H_0: p_{k1} = p_{k2} = \cdots = p_{kM} =:p_k \ (\forall k \in \{1, \ldots, K\})$,
  the test statistic $I_{(N)}$ approximately obeys the normal distribution with mean $\mu$ and variance $\sigma^2$, where
  \begin{align*}
    \mu &:= \frac{1}{(K-1) / 4}\left[ \sum_{k=1}^{K-1} (K-k) p_k\right],\\
    \sigma^2 &:= \frac{1}{nM (K-1)^2 / 16} \left[ \sum_{k=1}^{K-1} (K-k)^2 p_k (1 - p_k) - \sum_{1 \leq k, l \leq K-1 ; k \neq l} (K-k)(K-l) p_k p_l \right]. 
  \end{align*}
\end{theorem}

\begin{proof}
From the homogeneous condition $H_0$, we assume that
\begin{equation*}
  \left( X_{1m}, \ldots, X_{Km} \right) \sim
  \mathrm{MN} \left( n; p_1, \ldots, p_K \right) 
\end{equation*}
 for any laboratory $m$, where $\textrm{MN} (n; p_1, \ldots, p_k)$ denotes the multinomial distribution for $n$ trials each of which lies in $K$ categories with probabilities $p_1, \ldots, p_k$. 

 First, we approximate the multinominal distribution by a multi-normal distribution.
 Let $\mu_{K-1} := (np_1 , \ldots , np_{K-1})$ and $\Sigma_{K-1} := ( \sigma_{ij} )_{i, j = 1, \ldots, K-1},$ where
 \begin{equation*}
   \sigma_{ij} :=
        \begin{cases}
        n p_i (1 - p_i ) & \text{if} \ j = i,\\
        -n p_i p_j & \text{if} \ j \neq i.
     \end{cases}
    \end{equation*}
Then, for any laboratory $m$, any number of measurement results $n$ and $p_1, \ldots, p_K$, we can consider $(X_{1m}, \ldots, X_{Km})$ to be a vector of random variables that follows the multi-normal distribution $N( \mu_{K-1} , \Sigma_{K-1})$.

Next, we obtain
\begin{align*}
  I_{(N)}
  &= \hat{S}^2_{(B)} + \hat{h}^2_{(W)} + \frac{1}{(K-1) / 4} \sum_{k=1}^{K-1} \hat{\bar{F}}^2_k
  = \hat{h}^2_{(T)} + \frac{1}{(K-1) / 4} \sum_{k=1}^{K-1} \hat{\bar{F}}^2_k\\
  &= \frac{1}{(K-1) / 4} \sum_{k=1}^{K-1} \hat{\bar{F}}_k \left(1 - \hat{\bar{F}}_k\right)
    + \frac{1}{(K-1) / 4} \sum_{k=1}^{K-1} \hat{\bar{F}}^2_k
  = \frac{1}{(K-1) / 4} \sum_{k=1}^{K-1}  \hat{\bar{F}}_k \\
  &=  \frac{1}{M (K-1) /4} \sum_{k=1}^{K-1} \sum_{m=1}^M \sum_{l=1}^k \hat{p}_{km}
  =  \frac{1}{M (K-1) /4} \sum_{k=1}^{K-1} \sum_{m=1}^M \sum_{l=1}^k \frac{X_{li}}{n}\\
  &= \frac{1}{n M (K-1) /4} \sum_{m=1}^M
      \left(  X_{m1} + (X_{m1} + X_{m2}) + \cdots +
      (X_{m1} + \cdots + X_{m, K-1} ) \right)\\
  &= \frac{1}{n M (K-1) / 4}\sum_{m=1}^M \sum_{k=1}^{K-1} (K - k) X_{km}.
\end{align*}

Since $(X_{1m}, \ldots, X_{K-1, m}) \sim N( \mu_{K-1}, \Sigma_{K-1} )$ for any $m$,
we obtain
\begin{align*}
&  \sum_{k=1}^{K-1} (K - k) X_{km} \\
& \quad   \sim N \left(
    \sum_{k=1}^{K-1} (K - k) n p_k, \right.\\
& \quad \qquad \qquad \left. \sum_{k=1}^{K-1} (K-k)^2 np_k (1 - p_k)
    - \sum_{1 \leq k, l \leq K-1; k \neq l} (K-k) (K-l) n p_k p_l
                                                         \right)
\end{align*}
for any $m$.

Thus, the following holds:
\begin{align*}
  I_{(N)}
  &= \frac{1}{n M (K-1) /4} \sum_{m=1}^M \sum_{k=1}^{K-1} (K-k) X_{km}\\
  &\sim N  \left[
    \frac{1}{n (K-1) /4} \sum_{k=1}^{K-1} (K-k) n p_k, \right.\\
  & \qquad \qquad   M \left( \frac{1}{n M (K-1)/ 4} \right)^2 \times \\
  & \qquad \qquad \quad \left.  \left( \sum_{k=1}^{K-1} (K-k)^2 n p_k (1 - p_k) 
    - \sum_{1 \leq k, l \leq K-1 ; k \neq l} (K-k) (K-l) n p_k p_l\right)
    \right]\\
 &= N \left[ \frac{1}{(K-1) /4} \left( \sum_{k=1}^{K-1} (K-k) p_k \right), \right.\\
& \quad \left. \frac{1}{n M (K-1)^2 / 16} \left( \sum_{k=1}^{K-1} (K-k)^2 p_k (1-p_k) 
 - \sum_{1 \leq k, l \leq; k \neq l} (K-k)(K-l) p_k p_l \right) \right].
\end{align*}
This completes the proof. 
\end{proof}

\begin{remark}
  \label{rem:howtest}
  From Theorem~\ref{thm:main}, if there are no laboratory effects, then $I_{(N)}$ obeys the normal distribution.
  Therefore, by calculating $I_{(N)}$ using the results of a collaborative study on the measurement method we are interested in, and comparing it to the $\alpha$-percentile of the approximate distribution shown in Theorem~\ref{thm:main}, we can determine whether there are laboratory effects or not.
  More precisely, 
  \begin{enumerate}
  \item when $I_{(N)} > \alpha\text{-percentile of the approximate distribution}$, reject the homogeneous condition $H_0$; 
  \item when $I_{(N)} \leqq \alpha\text{-percentile of the approximate distribution}$, do not reject $H_0$.
    \end{enumerate}
    Since Theorem~\ref{thm:main} holds for an arbitrary number of categories $K$, our proposed statistical test, which is based on $I_{(N)}$, works for any number of categories $K$; while the statistical test based on $I_{(P)}$, which was introduced in Gadrich and Bashkansky~\cite{Gadrich.Bashkansky_2012_ORDANOVAAnalysis}, only works for $K=2$.
\end{remark}


  \subsection{Simulations on the two test statistics}
\label{sec:discus}

This section, only for the case of $K =3$, compares the two test statistics, $I_{(P)}$ and $I_{(N)}$, to several criteria.
Two criteria are used: (i) the constant number three and (ii) the upper fifth percentile of the exact distribution derived by a computer simulation.
The following two cases are treated: (a) $p_1 = p_2 = p_3 = 1/3$ and (b) $p_1 = 3/6, \ p_2 = 1/6, \ p_3=2/6$.
For each case, the number of laboratories $M$ is set to be $5$, $10$, and $20$ and the number of measured values in each laboratory $n$ is set to be $5$, $10$, and $20$. 

First, the following procedure was conducted to produce the exact distribution of $I_{(P)}$ and calculate the upper fifth percentile of $I_{(P)}$, which were performed using Mathematica 12.3~\cite{WolframResearchInc._2021_MathematicaVersion}.

\begin{enumerate}[label=(\roman*)]
\item $I^s_{(P)} := \emptyset.$
\item For each $m \in \{1, \ldots, M\}$, sample from a triple of random variables $(X_{1m}, X_{2m}, X_{3m})$ that follows the multi-nominal distribution $\textrm{MN}(n;p_1, p_2, p_3)$, say $(x_{1m}, x_{2m}, x_{3m})$.
\item Calculate the statistic $I_{(P)}$ using $(x_{1m}, x_{2m}, x_{3m})$.
  \item $I^s_{(P)} := I^s_{(P)} \cup I_{(P)}$.
\item Repeat procedures (ii)–(iv) $10,000$ times.
\item Calculate the value at the top of $5\%$ in the set of $I^s_{(P)}$, which is denoted by $I^s_{(P), 0.05}$.
\end{enumerate}

\begin{figure}[htbp]
  \centering \includegraphics[scale=0.7]{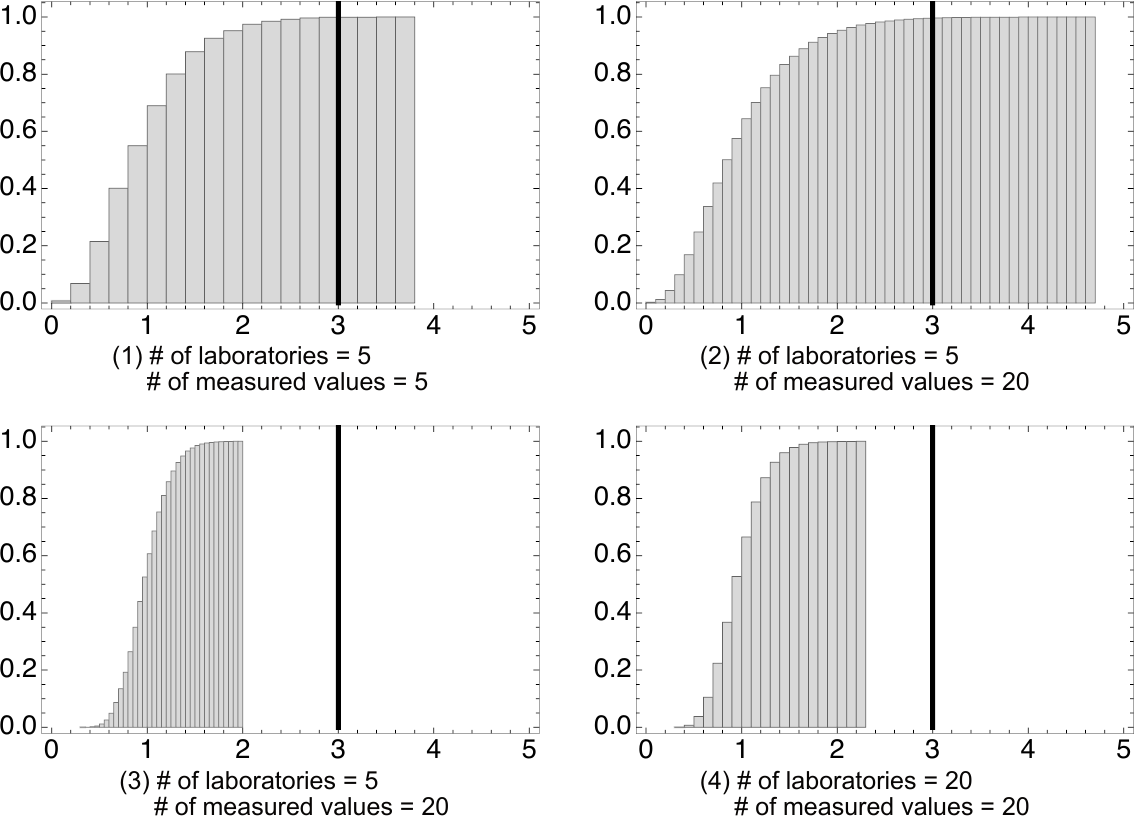}
  \caption{The light gray bars and the bold black lines show cumulative histograms of $I^s_{(P)}$ and the constant number three, respectively, in a part of the case (a). In each figure, the horizontal axis represents the values of $I^s_{(P)}$, whereas the vertical axis does the cumulative frequencies of $I^s_{(P)}$. The constant number three is much larger than the value at the top five percent in the set of $I^s_{(P)}$.} \label{fig:1}
\end{figure}

\begin{figure}[htbp]
  \centering
  \includegraphics[scale=0.7]{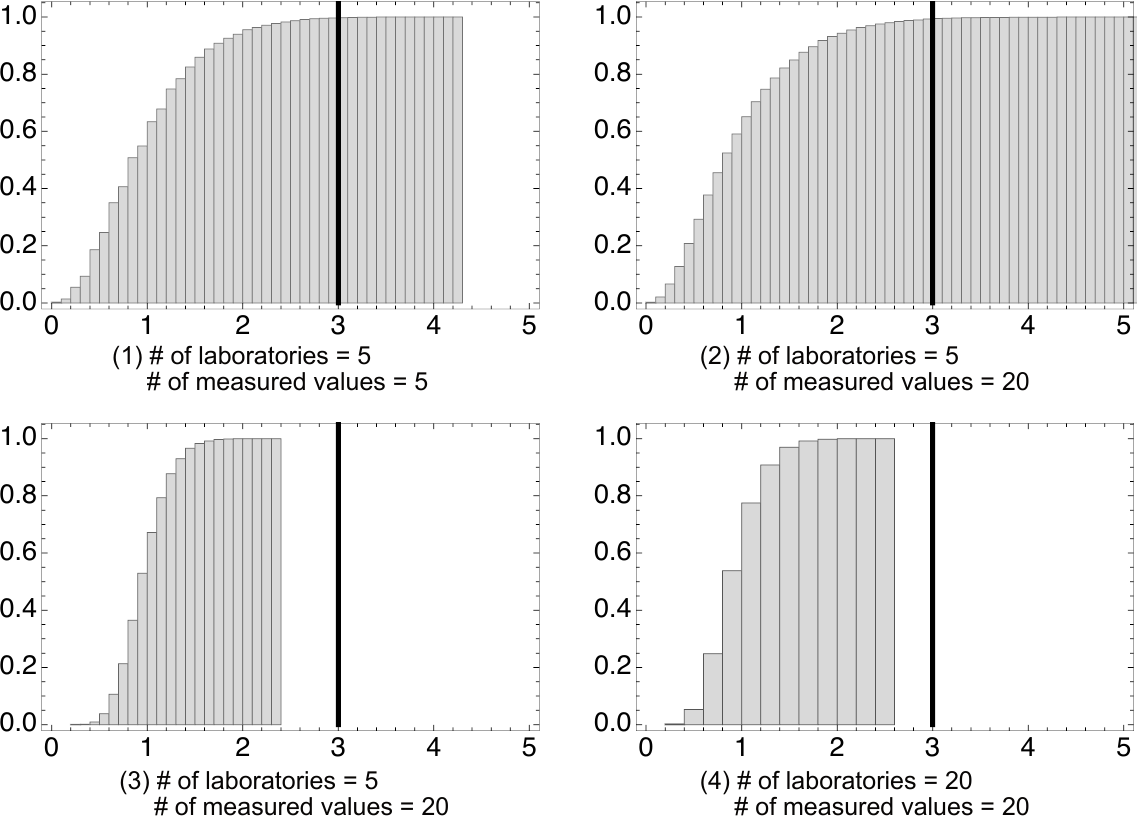}
  \caption{The light gray bars and the bold black lines show cumulative histograms of $I^s_{(P)}$ and the constant number three, respectively, in a part of the case (b).
    In each figure, the horizontal axis represents the values of $I^s_{(P)}$, whereas the vertical axis does the cumulative frequencies of $I^s_{(P)}$.
    The constant number three is much larger than the value at the top five percent in the set of $I^s_{(P)}$.} \label{fig:2}
\end{figure}

Figure~\ref{fig:1} presents a cumulative histogram of $I^s_{(P)}$ and the constant three.
In each plot, the horizontal and vertical axes represent, respectively, the values of the test statistics, shown as the light gray bars, and the cumulative probabilities, shown as the bold black line, for the cases where (1) $M=5$, $n=5$, (2) $M=5$, $n=20$,  (3) $M=20$, $n=5$, and  (4) $M=20$, $n=20$ of case (a).
Figure~\ref{fig:2} presents the same information as Figure~\ref{fig:1}, but for case (b).
If the $y$-axis values of the intersection points between the bold black lines and histograms equal approximately $0.95$,
then the criterion three, proposed in the previous study succeeds in detecting laboratory effects with a five-percent significance level. 

Table~\ref{tab:1} illustrates $I^s_{(P), 0.05}$, the relative error of the constant three for $I^s_{(P), 0.05}$, and the percentage of the area $\{I_{(P)} \in I^s_{(P)} \, | \, I_{(P)} \geq 3\}$ for case (a).
Table~\ref{tab:2} presents the same information as Table~\ref{tab:1}, but for case (b).
The constant three is much larger than $I^s_{(P), 0.05}$ for all cases.
In addition, the percentages of area  $\{I_{(P)} \in I^s_{(P)} \, | \, I_{(P)} \geq 3\}$ are considerably smaller than $I^s_{(P), 0.05}$ for all cases.
We therefore conclude that the constant three is too strict of a criterion for detecting laboratory effects when the number of categories is $K=3$.

\begin{table}[htbp]
  \caption{The upper $5$-percentiles of exact distribution by the simulation $I^s_{(P)}$, the relative errors of the constant number three for the upper $5$-percentiles, and the percentage of the area the area $\{I_{(P)} \in I^s_{(P)} \, | \, I_{(P)} \geq 3\}$ in the case (a) $p_1 = p_2 = p_3 = 1/2$.}
  {\begin{tabular}{cccccc}\hline
     & & Upper $5$- & Relative errors & The percentage\\
    \# of labs & \# of & percentile of $I^s_{(P)}$, & of $I^s_{(P), 0.05}$ for 3, & of  the area\\
     & repetition & $I^s_{(P), 0.05}$ & $|3 - I^s_{(P),0.05}| / I^s_{(P),0.05}$ & $\left\{ I_{(P)} \in I^s_{(P)} \, \right.$ \\
  &&&& $\qquad\quad \left. | \, I_{(P)} \geq 3 \right\}$ $(\% )$ \\\hline
    $5$ & $5$ &	$1.97$ & $0.52$ & $0.2$\\
    $5$ & $10$ & $2.00$ & $0.50$ & $0.3$\\
    $5$ &	$20$ & $2.07$ & $0.45$ & $0.5$\\
    $10$ & $5$ & $1.60$ & $0.87$ & $0.0$\\ 
    $10$ & $10$ & $1.65$ & $0.82$ & $0.0$\\
    $10$ & $20$ & $1.68$ & $0.78$ & $0.0$\\
    $20$ & $5$ & $1.41$ & $1.13$ & $0.0$\\
    $20$ & $10$ & $1.43$ & $1.10$ & $0.0$\\
    $20$ & $20$ & $1.46$ & $1.05$ & $0.0$\\ \hline
   \end{tabular}}
 \label{tab:1}
\end{table}

\begin{table}[htbp]
  \caption{The upper $5$-percentiles of exact distribution by the simulation $I^s_{(P)}$, the relative errors of the constant number three for the upper $5$-percentiles, and the percentage of the area the area $\{I_{(P)} \in I^s_{(P)} \, | \, I_{(P)} \geq 3\}$ in the case (b) $p_1 = 3/6$, $p_2 = 1/2$, $p_3 = 2/6$.}
 {\begin{tabular}{cccccc}\hline
     & & Upper $5$- &
    Relative errors &
    The percentage\\
    \# of labs & \# of & percentile of $I^s_{(P)}$, & of $I^s_{(P), 0.05}$ for 3, & of  the area\\
     & repetition & $I^s_{(P), 0.05}$ & $|3 - I^s_{(P),0.05}| / I^s_{(P),0.05}$ & $\left\{ I_{(P)} \in I^s_{(P)} \, \right.$ \\
  &&&& $\qquad\quad \left. | \, I_{(P)} \geq 3 \right\}$ $(\% )$ \\\hline
    $5$ & $5$ &	$2.07$ & $0.45$ & $0.3$\\
    $5$ & $10$ & $2.13$ & $0.41$ & $0.9$\\
    $5$ &	$20$ & $2.15$ & $0.39$ & $0.8$\\
    $10$ & $5$ & $1.68$ & $0.79$ & $0.0$\\ 
    $10$ & $10$ & $1.71$ & $0.75$ & $0.0$\\
    $10$ & $20$ & $1.74$ & $0.73$ & $0.1$\\
    $20$ & $5$ & $1.44$ & $1.08$ & $0.0$\\
    $20$ & $10$ & $1.46$ & $1.05$ & $0.0$\\
    $20$ & $20$ & $1.52$ & $0.98$
                    & $0.0$\\\hline
  \end{tabular}}
\label{tab:2}
\end{table}

Next, this subsection compares the exact distribution of $I_{(N)}$ derived by a computer simulation and the approximate distribution proposed in the previous subsection for the same conditions as above.
The following procedure was conducted to produce the exact distribution of $I_{(N)}$, which were performed using Mathematica 12.3~\cite{WolframResearchInc._2021_MathematicaVersion}.
\begin{enumerate}[label=(\roman*)]
\item $I^s_{(N)} := \emptyset.$
\item For each $m \in \{1, \ldots, M\}$, sample from a triple of random variables $(X_{1m}, X_{2m}, X_{3m})$  that follows the multinominal distribution $\textrm{MN}(n;p_1, p_2, p_3)$, say $(x_{1m}, x_{2m}, x_{3m})$.
\item Calculate the statistic $I_{(N)}$ using $(x_{1m}, x_{2m}, x_{3m})$.
\item $I^s_{(N)} := I^s_{(N)} \cup I_{(N)}.$
\item Repeat procedures (ii)–(iv) $10,000$ times.
\end{enumerate}
The approximate distribution $I_{(N)}$ is calculated using Theorem~\ref{thm:main}, and it is denoted by $I_{(N)}^a$. 

\begin{figure}[htbp]
  \centering
  \includegraphics[scale=0.7]{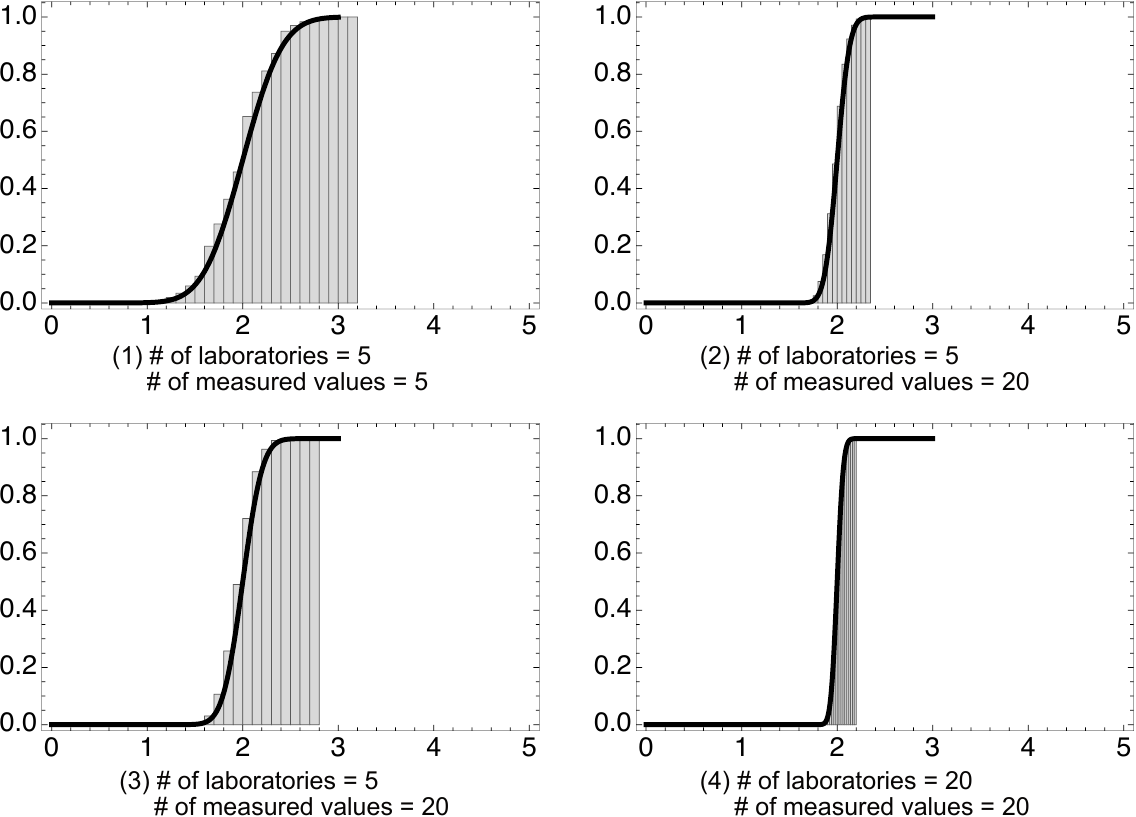}
  \caption{The light gray bars are cumulative histograms of $I^s_{(N)}$, whereas the bold black lines represent the distribution functions of $I^a_{(N)}$ in a part of case(a).
   In each plot, the horizontal axis represents the values of the test statistics, whereas the vertical axis represents the cumulative probabilities.
    The cumulative probabilities and the values of the distribution function of $I^a_{(N)}$ are almost identical in all plots.}\label{fig:3}
\end{figure}

\begin{figure}[htbp]
  \centering
  \includegraphics[scale=0.7]{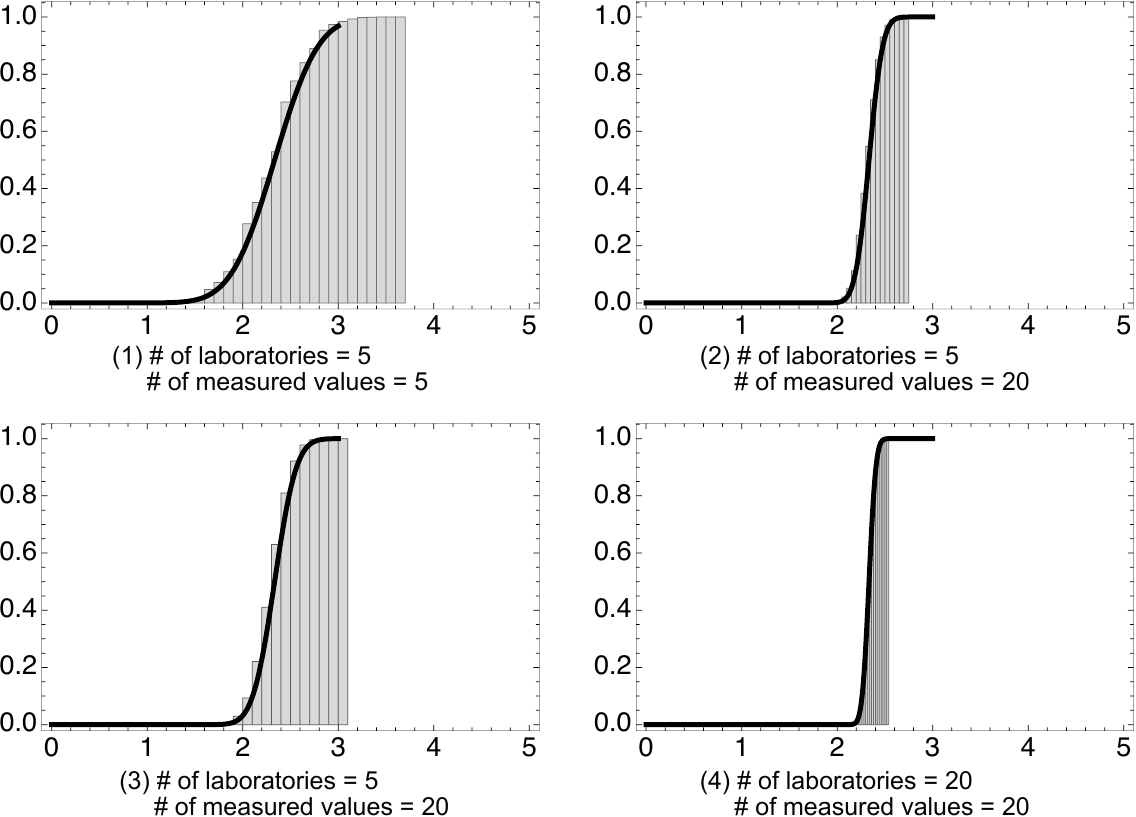}
  \caption{The light gray bars are cumulative histograms of $I^s_{(N)}$, whereas the bold black lines represent the distribution functions of $I^a_{(N)}$ in a part of case(b).
   In each plot, the horizontal axis represents the values of the test statistics, whereas the vertical axis represents the cumulative probabilities.
    The cumulative probabilities and the values of the distribution function of $I^a_{(N)}$ are almost identical in all plots.}\label{fig:4}
\end{figure}

Figure~\ref{fig:3} illustrates both cumulative histograms of $I_{(N)}^s$ and the distribution function of $I_{(N)}^a$ for the cases where (1) $M=5$, $n=5$, (2) $M=5$, $n=20$,  (3) $M=20$, $n=5$, and  (4) $M=20$, $n=20$ of case (a).
In each figure, the horizontal axis represents the values of either $I_{(N)}^s$ or $I_{(N)}^a$, and the vertical axis does these cumulative frequency.
The light gray bars and bold black line represent the cumulative frequency of $I_{(N)}^s$ and the distribution function of $I_{(N)}^a$, respectively.
Figure~\ref{fig:4} presents the same information as Figure~\ref{fig:3}, but for case (b).
Only several differences exist between $I_{(N)}^s$ and $I_{(N)}^a$;
therefore, $I_{(N)}^a$ sufficiently approximates the test statistic $I_{(N)}$.
In addition, Table~\ref{tab:5} illustrates the upper fifth percentiles of $I_{(N)}^s$ and  $I_{(N)}^a$,
and the relative errors of $I_{(N)}^a$ for the upper fifth percentile in case (a).
Table~\ref{tab:6} illustrates the same information as Table~\ref{tab:5}, but for case (b).
It can be seen that the relative errors decrease when the number of laboratories and measured values increases.
However, the values of the upper fifth percentiles are almost identical between $I_{(N)}^s$ and $I_{(N)}^a$, and the relative errors of $I_{(N)}^a$ for $I_{(N)}^s$ are less than $0.03$.
Therefore, we conclude that the approximate continuous distribution $I_{(N)}^a$ can be used to conduct statistical tests for detecting laboratory effects.
To detect laboratory effects with a $5 \%$-significance level, the upper fifth percentile of the normal distribution $I_{(N)}^a$ should be used.

\begin{table}[htbp]
  \caption{Upper fifth percentiles of the exact distribution by simulation, $I^s_{(N)}$, and the approximate distribution, $I^a_{(N)}$ and the relative error of $I^a_{(N)}$ for $I^s_{(N)}$ for case (a) $p_1 = p_2 = p_3 = 1/3$.
    The upper fifth percentile of $I^s_{(N)}$ and that of $I^a_{(N)}$ are almost identical in each number of laboratories and measurement results.}
  {\begin{tabular}{cccccc}\hline
     & \# of & \multicolumn{2}{c}{Upper fifth percentile} & Relative errors of of $I^a_{(N)}$ for $I^s_{(N)}$,\\
    \# of labs & repetition & $I^s_{(N)}$ & $I^a_{(N)}$ & $|I^a_{(N)} - I^s_{(N)}| / I^s_{(N)}$\\\hline
    $5$ & $5$ &	$2.56$ & $2.54$ & $0.01$\\
    $5$ & $10$ & $2.36$ & $2.38$ & $0.01$\\
    $5$ &	$20$ & $2.17$ & $2.17$ & $0.00$\\
    $10$ & $5$ & $2.36$ & $2.38$ & $0.01$\\ 
    $10$ & $10$ & $2.26$ & $2.27$ & $0.0$\\
    $10$ & $20$ & $2.18$ & $2.19$ & $0.00$\\
    $20$ & $5$ & $2.26$ & $2.27$ & $0.00$\\
    $20$ & $10$ & $2.18$ & $2.19$ & $0.00$\\
    $20$ & $20$ & $2.09$ & $2.08$
                    & $0.00$\\\hline
  \end{tabular}
} \label{tab:5}
\end{table}

\begin{table}[htbp]
  \caption{Upper fifth percentiles of the exact distribution by simulation, $I^s_{(N)}$, and the approximate distribution, $I^a_{(N)}$ and the relative error of $I^a_{(N)}$ for $I^s_{(N)}$ for case (b) $p_1 = 3/6$, $p_2 = 1/2$, $p_3 = 2/6$.
    The upper fifth percentile of $I^s_{(N)}$ and that of $I^a_{(N)}$ are almost identical in each number of laboratories and measurement results.}
  {\begin{tabular}{cccccc}\hline
    & \# of & \multicolumn{2}{c}{Upper fifth percentile} & Relative errors of of $I^a_{(N)}$ for $I^s_{(N)}$,\\
    \# of labs & repetition & $I^s_{(N)}$ & $I^a_{(N)}$ & $|I^a_{(N)} - I^s_{(N)}| / I^s_{(N)}$\\\hline
    $5$ & $5$ &	$2.88$ & $2.92$ & $0.02$\\
    $5$ & $10$ & $2.72$ & $2.75$ & $0.01$\\
    $5$ &	$20$ & $2.52$ & $2.52$ & $0.00$\\
    $10$ & $5$ & $2.72$ & $2.75$ & $0.01$\\ 
    $10$ & $10$ & $2.64$ & $2.63$ & $0.00$\\
    $10$ & $20$ & $2.46$ & $2.47$ & $0.00$\\
    $20$ & $5$ & $2.64$ & $2.63$ & $0.00$\\
    $20$ & $10$ & $2.54$ & $2.54$ & $0.00$\\
    $20$ & $20$ & $2.43$ & $2.43$
               & $0.00$\\\hline
  \end{tabular}
  }\label{tab:6}
\end{table}

\section{Real-world example}
\label{sec:ex}

This section analyzes the measured values of an interlaboratory comparison study reported in AIST~\cite{AIST_2018_AnnualReport}.
The study was conducted to assess the precesion of intratracheal administration testing~\cite{Driscoll.Costa.ea_2000_IntratrachealInstillation}, which is an \textit{in vivo} screening method for evaluating the pulmonary toxicity of nanomaterials~\cite{Nakanishi.Morimoto.ea_2015_RiskAssessment}.
The study consisted of five laboratories.
 Each laboratory examined 19 pathological findings using five rats and reported one of the following scores for each finding and each rat: $-, \pm , +, ++$ and $+++$.
Here score $-$ indicates that the rat had no effect on the focused findings, whereas scores $\pm, +, ++$  and $+++$ indicate that the rat did have an effect on the focused findings.
Scores $\pm$ and $+++$ represent the weakest and strongest effects, respectively. 

As examples, the present study addresses two of the 19 pathological findings and analyses the obtained data after scores $-, \pm, +, ++$ and $+++$ are converted to category numbers 1, 2, 3, 4 and 5, respectively.

Since $K=5$, $M=5$, and $n=5$, the proposed test statistic $I_{(N)}$ becomes
\begin{equation*}
 I_{(N)} = \frac{1}{25} \sum_{m=1}^5 \sum_{k=1}^4 (5-k) X_{km},
\end{equation*}
and, under the homogeneous condition, $H_0 : p_{k1} = p_{k1} = \cdots = p_{k5}:= p_k  \ (\forall k \in \{1, \ldots, 5\})$,
\begin{align*}
 I_{(N)} & \sim \\
 & N \left[ \sum_{k=1}^4 (5-k) p_k,
   \frac{1}{25} \left\{ \sum_{k=1}^4 (5-k)^2 p_k (1- p_k) - \sum_{1 \leq k,l \leq 4; k \neq l} (5-k) (5-l) p_k p_l \right\} \right].
 \end{align*}

Because the value of $p_k$ is unknown in real-world examples, the estimate of $p_k$, $\hat{\bar{p}}_k$, is used instead; see Remark~\ref{rem:unknown_param} for this replacement.
Table~\ref{tab:3} presents the measured values on the appearance of alveolar macrophages following the administration of $0.13 \, \mathrm{mg/kg}$ weight of a multiwall carbon nanotube (MWCNT).
 In this case, $I_{(N)} = 1.80$ and the upper fifth percentile of the normal distribution is $2.05$.
 Since $I_{(N)} < 2.05$, this demonstrates that we cannot say there is a laboratory effect. 
 This statistical test result imply that the intratracheal administration testing does not differ among laboratories for the  pathological finding.

  \begin{table}[htbp]
   \caption{Results of a collaborative study on the appearance of alveolar macrophages following the administration of $0.13$ mg/kg weight of MWCNT, reported in AIST~\cite{AIST_2018_AnnualReport}.
     The nonnegative integer in each cell stands for the number of rats belonging to each laboratory and each score.}
   {\begin{tabular}{cccccc} \hline
     & \multicolumn{5}{c}{\# of rats reported in each category}\\
     Lab   & \quad $1$ \quad & \quad $2$ \quad & \quad $3$ \quad & \quad $4$ \quad & \quad $5$ \quad \\\hline
     Lab A & $0$ & $0$ & $0$ & $5$ & $0$\\
     Lab B & $0$ & $0$ & $1$ & $4$ & $0$\\
     Lab C & $0$ & $3$ & $2$ & $0$ & $0$\\
     Lab D & $0$ & $0$ & $5$ & $0$ & $0$\\
     Lab E & $0$ & $2$ & $2$ & $1$ & $0$\\ \hline
   \end{tabular}}
 \label{tab:3}
 \end{table}

 \begin{remark}
   \label{rem:unknown_param}
   In general, it is necessary to be careful when replacing unknown parameters with their estimates.
   However, since $\hat{\bar{p}}_k$ is a consistent estimator of $p_k$, $\hat{\bar{p}}_k$ is considered to be almost identical with $p_k$ when the number of samples is large.
   Indeed, this replacement is often applied when confidence intervals for binomial or multinomial proportion are derived.
   In the example, the number of samples is $25$, but if one considers that this sample size is too small and wants to deal strictly with the unknown parameter, further research is needed. 
   To use upper or lower confidence limits is a possible method, instead of the point estimate of p, but there are several methods for computing simultaneous confidence intervals for multinomial proportions, and which of them is most appropriate has been discussed; see, e.g., \cite{Jin_2013_Computingexact} and \cite{Batterton.Schubert.ea_2021_Exactconfidence}.
 \end{remark}
 
\section{Concluding remarks}
\label{sec:conc}

The present study gave an approximate multinominal distribution for the test statistic in ORDANOVA for an arbitrary number of ordinal-level measured data.
Using the $\alpha$-percentile of the approximate distribution, statistical tests can be conducted to detect laboratory effects.
Moreover, we discussed the accuracy of the results of the present study and previous studies.
The discussion in Section~\ref{sec:discus} suggests that the $\alpha$-percentile of the approximate multinormal distribution in the present study is effective for conducting statistical tests to detect laboratory effects, even if the number of measured values at each laboratory and that of laboratories are not large.
In addition, the present study analyzed the obtained measured values of a collaborative study on intratracheal administration testing as a real-world example.
Thus, the results of the present study would be make the ORDANOVA more powerful and practicable, because the previous study introduced an approximate distribution of its proposed test statistic only for the two-level ordinal, that is, binary measured data.

However, the approximation accuracy would not be so high since the proposed methodology was based on the normal approximation of multinomial distributions.
Therefore, it may be possible to construct a better statistical method for detecting laboratory effects on ORDANOVA by using a better approximation.

Also, the present study focused solely on ORDANOVA for analyzing the ordinal measured values obtained from collaborative studies.
However, methods other than ORDANOVA have been proposed for ordinal or binary data.
For example, Wilrich~\cite{Wilrich_2010_DeterminationPrecision} has proposed an ISO 5725-based method, van Wieringen \& de Mast~\cite{vanWieringen.deMast_2008_MeasurmentSystem} has developed a Gauge R\&R-based method, and Langton \textit{et al.}~\cite{Scotter.Langton.ea_2001_ValidationISO} has introduced some new concepts for qualitative measured values.
However, these studies only address the case of two-level categories.
Hence, further studies are necessary to extend these methods applicable to three- or more-level categories, and can be subsequently compared to our proposed method.

\section*{Acknowledgements}
J.T. and T.S. are supported by JSPS KAKENSHI Grant Numbers JP20K12203, JP16K21674 and JP15K01207.
Also, the present study is supported by ``Survey on standardisation of intratracheal administration study for nanomaterials related issues'' funded by the Ministry of Economy, Trade and Industry (METI) of Japan.

\section*{declaration of interest statement}
All the authors have no relevant interest(s) to disclose.

\bibliographystyle{unsrt}
\bibliography{TestStatsORDANOVA_ref}  

%
%
%
%

\end{document}